\documentclass[aps,prl,twocolumn,showpacs,superscriptaddress,floatfix]{revtex4}

\usepackage{epsfig}

\begin{document}


\title{ Energy and centrality dependence of $\bar{p}$ and $p$ production
            and the $\bar{\Lambda}$/$\bar{p}$ ratio \\
 
   in Pb+Pb collisions between 20$A$ GeV and 158$A$ GeV }

\vspace{0.5cm}



\affiliation{NIKHEF, Amsterdam, Netherlands.}  
\affiliation{Department of Physics, University of Athens, Athens, Greece.}
\affiliation{Comenius University, Bratislava, Slovakia.}
\affiliation{KFKI Research Institute for Particle and Nuclear Physics,
             Budapest, Hungary.}
\affiliation{MIT, Cambridge, USA.}
\affiliation{Institute of Nuclear Physics, Cracow, Poland.}
\affiliation{Gesellschaft f\"{u}r Schwerionenforschung (GSI),
             Darmstadt, Germany.} 
\affiliation{Joint Institute for Nuclear Research, Dubna, Russia.}
\affiliation{Fachbereich Physik der Universit\"{a}t, Frankfurt, Germany.}
\affiliation{CERN, Geneva, Switzerland.}
\affiliation{Institute of Physics \'Swi{\,e}tokrzyska Academy, Kielce, Poland.}
\affiliation{Fachbereich Physik der Universit\"{a}t, Marburg, Germany.}
\affiliation{Max-Planck-Institut f\"{u}r Physik, Munich, Germany.}
\affiliation{Institute of Particle and Nuclear Physics, Charles
             University, Prague, Czech Republic.}
\affiliation{Department of Physics, Pusan National University, Pusan,
             Republic of Korea.} 
\affiliation{Nuclear Physics Laboratory, University of Washington,
             Seattle, WA, USA.} 
\affiliation{Atomic Physics Department, Sofia University St.~Kliment
             Ohridski, Sofia, Bulgaria.} 
\affiliation{Institute for Nuclear Research and Nuclear Energy, Sofia, Bulgaria.} 
\affiliation{Institute for Nuclear Studies, Warsaw, Poland.}
\affiliation{Institute for Experimental Physics, University of Warsaw,
             Warsaw, Poland.} 
\affiliation{Rudjer Boskovic Institute, Zagreb, Croatia.}


\author{C.~Alt}
\affiliation{Fachbereich Physik der Universit\"{a}t, Frankfurt, Germany.}
\author{T.~Anticic} 
\affiliation{Rudjer Boskovic Institute, Zagreb, Croatia.}
\author{B.~Baatar}
\affiliation{Joint Institute for Nuclear Research, Dubna, Russia.}
\author{D.~Barna}
\affiliation{KFKI Research Institute for Particle and Nuclear Physics,
             Budapest, Hungary.} 
\author{J.~Bartke}
\affiliation{Institute of Nuclear Physics, Cracow, Poland.}
\author{L.~Betev}
\affiliation{CERN, Geneva, Switzerland.}
\author{H.~Bia{\l}\-kowska} 
\affiliation{Institute for Nuclear Studies, Warsaw, Poland.}
\author{C.~Blume}
\affiliation{Fachbereich Physik der Universit\"{a}t, Frankfurt, Germany.}
\author{B.~Boimska}
\affiliation{Institute for Nuclear Studies, Warsaw, Poland.}
\author{M.~Botje}
\affiliation{NIKHEF, Amsterdam, Netherlands.}
\author{J.~Bracinik}
\affiliation{Comenius University, Bratislava, Slovakia.}
\author{R.~Bramm}
\affiliation{Fachbereich Physik der Universit\"{a}t, Frankfurt, Germany.}
\author{P.~Bun\v{c}i\'{c}}
\affiliation{CERN, Geneva, Switzerland.}
\author{V.~Cerny}
\affiliation{Comenius University, Bratislava, Slovakia.}
\author{P.~Christakoglou}
\affiliation{Department of Physics, University of Athens, Athens, Greece.}
\author{O.~Chvala}
\affiliation{Institute of Particle and Nuclear Physics, Charles
             University, Prague, Czech Republic.} 
\author{J.G.~Cramer}
\affiliation{Nuclear Physics Laboratory, University of Washington,
             Seattle, WA, USA.} 
\author{P.~Csat\'{o}} 
\affiliation{KFKI Research Institute for Particle and Nuclear Physics,
             Budapest, Hungary.}
\author{P.~Dinkelaker}
\affiliation{Fachbereich Physik der Universit\"{a}t, Frankfurt, Germany.}
\author{V.~Eckardt}
\affiliation{Max-Planck-Institut f\"{u}r Physik, Munich, Germany.}
\author{D.~Flierl}
\affiliation{Fachbereich Physik der Universit\"{a}t, Frankfurt, Germany.}
\author{Z.~Fodor}
\affiliation{KFKI Research Institute for Particle and Nuclear Physics,
             Budapest, Hungary.} 
\author{P.~Foka}
\affiliation{Gesellschaft f\"{u}r Schwerionenforschung (GSI),
             Darmstadt, Germany.} 
\author{V.~Friese}
\affiliation{Gesellschaft f\"{u}r Schwerionenforschung (GSI),
             Darmstadt, Germany.} 
\author{J.~G\'{a}l}
\affiliation{KFKI Research Institute for Particle and Nuclear Physics,
             Budapest, Hungary.} 
\author{M.~Ga\'zdzicki}
\affiliation{Fachbereich Physik der Universit\"{a}t, Frankfurt, Germany.}
\affiliation{Institute of Physics \'Swi{\,e}tokrzyska Academy, Kielce, Poland.}
\author{V.~Genchev}
\affiliation{Institute for Nuclear Research and Nuclear Energy, Sofia, Bulgaria.}
\author{G.~Georgopoulos}
\affiliation{Department of Physics, University of Athens, Athens, Greece.}
\author{E.~G{\l}adysz}
\affiliation{Institute of Nuclear Physics, Cracow, Poland.}
\author{K.~Grebieszkow}
\affiliation{Institute for Experimental Physics, University of Warsaw,
             Warsaw, Poland.} 
\author{S.~Hegyi}
\affiliation{KFKI Research Institute for Particle and Nuclear Physics,
             Budapest, Hungary.} 
\author{C.~H\"{o}hne}
\affiliation{Gesellschaft f\"{u}r Schwerionenforschung (GSI),
             Darmstadt, Germany.}
\author{K.~Kadija}
\affiliation{Rudjer Boskovic Institute, Zagreb, Croatia.}
\author{A.~Karev}
\affiliation{Max-Planck-Institut f\"{u}r Physik, Munich, Germany.}
\author{M.~Kliemant}
\affiliation{Fachbereich Physik der Universit\"{a}t, Frankfurt, Germany.}
\author{S.~Kniege}
\affiliation{Fachbereich Physik der Universit\"{a}t, Frankfurt, Germany.}
\author{V.I.~Kolesnikov}
\affiliation{Joint Institute for Nuclear Research, Dubna, Russia.}
\author{E.~Kornas}
\affiliation{Institute of Nuclear Physics, Cracow, Poland.}
\author{R.~Korus}
\affiliation{Institute of Physics \'Swi{\,e}tokrzyska Academy, Kielce, Poland.}
\author{M.~Kowalski}
\affiliation{Institute of Nuclear Physics, Cracow, Poland.}
\author{I.~Kraus}
\affiliation{Gesellschaft f\"{u}r Schwerionenforschung (GSI),
             Darmstadt, Germany.} 
\author{M.~Kreps}
\affiliation{Comenius University, Bratislava, Slovakia.}
\author{A.~Laslo}
\affiliation{KFKI Research Institute for Particle and Nuclear Physics,
             Budapest, Hungary.}
\author{M.~van~Leeuwen}
\affiliation{NIKHEF, Amsterdam, Netherlands.}
\author{P.~L\'{e}vai}
\affiliation{KFKI Research Institute for Particle and Nuclear Physics,
             Budapest, Hungary.} 
\author{L.~Litov}
\affiliation{Atomic Physics Department, Sofia University St.~Kliment
             Ohridski, Sofia, Bulgaria.} 
\author{B.~Lungwitz}
\affiliation{Fachbereich Physik der Universit\"{a}t, Frankfurt, Germany.}
\author{M.~Makariev}
\affiliation{Atomic Physics Department, Sofia University St.~Kliment
             Ohridski, Sofia, Bulgaria.} 
\author{A.I.~Malakhov}
\affiliation{Joint Institute for Nuclear Research, Dubna, Russia.}
\author{M.~Mateev}
\affiliation{Atomic Physics Department, Sofia University St.~Kliment
             Ohridski, Sofia, Bulgaria.} 
\author{G.L.~Melkumov}
\affiliation{Joint Institute for Nuclear Research, Dubna, Russia.}
\author{A.~Mischke}
\affiliation{NIKHEF, Amsterdam, Netherlands.} 
\author{M.~Mitrovski}
\affiliation{Fachbereich Physik der Universit\"{a}t, Frankfurt, Germany.}
\author{J.~Moln\'{a}r}
\affiliation{KFKI Research Institute for Particle and Nuclear Physics,
             Budapest, Hungary.} 
\author{St.~Mr\'owczy\'nski}
\affiliation{Institute of Physics \'Swi{\,e}tokrzyska Academy, Kielce, Poland.}
\author{V.~Nicolic}
\affiliation{Rudjer Boskovic Institute, Zagreb, Croatia.} 
\author{G.~P\'{a}lla}
\affiliation{KFKI Research Institute for Particle and Nuclear Physics,
             Budapest, Hungary.} 
\author{A.D.~Panagiotou}
\affiliation{Department of Physics, University of Athens, Athens, Greece.}
\author{D.~Panayotov}
\affiliation{Atomic Physics Department, Sofia University St.~Kliment
             Ohridski, Sofia, Bulgaria.} 
\author{A.~Petridis}
\affiliation{Department of Physics, University of Athens, Athens, Greece.}
\author{M.~Pikna}
\affiliation{Comenius University, Bratislava, Slovakia.}
\author{D.~Prindle}
\affiliation{Nuclear Physics Laboratory, University of Washington,
             Seattle, WA, USA.}
\author{F.~P\"{u}hlhofer}
\affiliation{Fachbereich Physik der Universit\"{a}t, Marburg, Germany.}
\author{R.~Renfordt}
\affiliation{Fachbereich Physik der Universit\"{a}t, Frankfurt, Germany.}
\author{C.~Roland}
\affiliation{MIT, Cambridge, USA.}
\author{G.~Roland}
\affiliation{MIT, Cambridge, USA.}
\author{M.~Rybczy\'nski}
\affiliation{Institute of Physics \'Swi{\,e}tokrzyska Academy, Kielce, Poland.}
\author{A.~Rybicki}
\affiliation{Institute of Nuclear Physics, Cracow, Poland.}
\affiliation{CERN, Geneva, Switzerland.}
\author{A.~Sandoval}
\affiliation{Gesellschaft f\"{u}r Schwerionenforschung (GSI),
             Darmstadt, Germany.} 
\author{N.~Schmitz}
\affiliation{Max-Planck-Institut f\"{u}r Physik, Munich, Germany.}
\author{T.~Schuster}
\affiliation{Fachbereich Physik der Universit\"{a}t, Frankfurt, Germany.}
\author{P.~Seyboth}
\affiliation{Max-Planck-Institut f\"{u}r Physik, Munich, Germany.}
\author{F.~Sikl\'{e}r}
\affiliation{KFKI Research Institute for Particle and Nuclear Physics,
             Budapest, Hungary.} 
\author{B.~Sitar}
\affiliation{Comenius University, Bratislava, Slovakia.}
\author{E.~Skrzypczak}
\affiliation{Institute for Experimental Physics, University of Warsaw,
             Warsaw, Poland.} 
\author{G.~Stefanek}
\affiliation{Institute of Physics \'Swi{\,e}tokrzyska Academy, Kielce, Poland.}
\author{R.~Stock}
\affiliation{Fachbereich Physik der Universit\"{a}t, Frankfurt, Germany.}
\author{C.~Strabel}
\affiliation{Fachbereich Physik der Universit\"{a}t, Frankfurt, Germany.}
\author{H.~Str\"{o}bele}
\affiliation{Fachbereich Physik der Universit\"{a}t, Frankfurt, Germany.}
\author{T.~Susa}
\affiliation{Rudjer Boskovic Institute, Zagreb, Croatia.}
\author{I.~Szentp\'{e}tery}
\affiliation{KFKI Research Institute for Particle and Nuclear Physics,
             Budapest, Hungary.} 
\author{J.~Sziklai}
\affiliation{KFKI Research Institute for Particle and Nuclear Physics,
             Budapest, Hungary.} 
\author{P.~Szymanski}
\affiliation{CERN, Geneva, Switzerland.} 
\affiliation{Institute for Nuclear Studies, Warsaw, Poland.}
\author{V.~Trubnikov} 
\affiliation{Institute for Nuclear Studies, Warsaw, Poland.}
\author{D.~Varga}
\affiliation{KFKI Research Institute for Particle and Nuclear Physics,
             Budapest, Hungary.} 
\affiliation{CERN, Geneva, Switzerland.}
\author{M.~Vassiliou}
\affiliation{Department of Physics, University of Athens, Athens, Greece.}
\author{G.I.~Veres}
\affiliation{KFKI Research Institute for Particle and Nuclear Physics,
             Budapest, Hungary.} 
\affiliation{MIT, Cambridge, USA.}
\author{G.~Vesztergombi}
\affiliation{KFKI Research Institute for Particle and Nuclear Physics,
             Budapest, Hungary.}
\author{D.~Vrani\'{c}}
\affiliation{Gesellschaft f\"{u}r Schwerionenforschung (GSI),
             Darmstadt, Germany.} 
\author{A.~Wetzler}
\affiliation{Fachbereich Physik der Universit\"{a}t, Frankfurt, Germany.}
\author{Z.~W{\l}odarczyk}
\affiliation{Institute of Physics \'Swi{\,e}tokrzyska Academy, Kielce, Poland.}
\author{I.K.~Yoo}
\affiliation{Department of Physics, Pusan National University, Pusan,
             Republic of Korea.} 
\author{J.~Zim\'{a}nyi}
\affiliation{KFKI Research Institute for Particle and Nuclear Physics,
             Budapest, Hungary.} 


\collaboration{The NA49 collaboration}
\noaffiliation


\begin{abstract}
 The transverse mass ($m_t$) distributions for antiprotons are 
 measured at midrapidity 
for minimum bias Pb+Pb collisions at 158$A$ GeV and for central Pb+Pb 
collisions at 20, 30, 40 and 80 $A$ GeV beam energies in the fixed target 
experiment NA49 at the CERN SPS.
The rapidity density $dn/dy$, inverse slope parameter $T$ and mean transverse 
mass $\langle m_{t}\rangle$ derived from the $m_t$ distributions are studied as 
a function of the incident energy and the collision centrality and compared 
to the relevant data on proton production.
The shapes of the $m_t$ distributions of $\bar{p}$ and $p$ are very similar.
The ratios of the particle yields, $\bar{p}$/$p$ and $\bar{\Lambda}$/$\bar{p}$,
are also analysed. The $\bar{p}$/$p$ ratio exhibits an increase with decreasing 
centrality and a steep rise with increasing beam energy.
The $\bar{\Lambda}$/$\bar{p}$ ratio increases beyond unity with decreasing beam 
energy.

\end{abstract}

\pacs{xx}

\maketitle

\section{Introduction}
Copious antibaryon as well as strangeness production in relativistic heavy 
ion collisions relative to the corresponding yields observed in elementary 
hadronic interactions have been suggested as signatures of the QCD phase
transition to a deconfined partonic state, the quark-gluon plasma \cite{b1,b2}. 
The enhancement was expected to arise from gluon fragmentation 
into quark-antiquark pairs which is believed to have a significantly lower  
threshold than baryon-antibaryon and strange-antistrange hadron pair
production channels.
More generally, such enhancements should be a consequence of the creation of 
a large volume of high energy density matter, uniquely characteristic of
central relativistic nuclear collisions \cite{b2}.

Meanwhile, in the net-baryon rich systems, significant annihilation losses 
could occur before antibaryons escape the collision volume.
Thus a systematic study of antiproton production has been proposed as an 
indirect way of measuring the baryon density created at the instant of 
hadron formation \cite{b3}. The latter is related to the degree of baryon 
stopping achieved during the early interpenetration phase of the collision. 
It has also been pointed out that antiprotons can be regenerated in multi-meson 
reactions during the final hadron gas stage \cite{b4,b5}.
Experimentally, the interplay between baryon stopping,
baryon pair production 
and antibaryon annihilation can be assessed through the measurement of 
antiproton and proton yields, and the yield ratios. This is the subject
of the study presented in this paper for the CERN SPS energy range.

As no antibaryons are contained in the initial nuclear projectiles their yields 
and spectra are determined predominantly by processes occurring subsequent 
to the primordial baryon stopping mechanisms. Of particular interest is the 
ratio of antilambda ($\bar{\Lambda}$) to antiproton ($\bar{p}$) which, 
intuitively, reflects the relative abundances of anti-strange to anti-light
quarks at the stage of formation of the finally observed antibaryons.
The $\bar{\Lambda}/\bar{p}$ production ratio amounts to about 0.25 in 
elementary hadron collisions at SPS energy \cite{b6}, but it has been 
found to significantly exceed unity in previous studies of central 
nucleus-nucleus collisions at AGS energies \cite{b7,b8,b9}. 
From top SPS energy ($\sqrt{s_{_{NN}}}$=17.3 GeV) to RHIC energy 
($\sqrt{s_{_{NN}}}$=130 GeV) 
this ratio is known to stay at about unity \cite{b10,b11}, a value also obtained as 
an upper limit (reflecting the general observation on strangeness enhancement 
in central A+A collisions) in recent theoretical studies employing statistical, 
and microscopic transport models \cite{b12,b13,b14,b15,b16}. 
As, in particular, the latter dynamical models attempt to incorporate both
the effects of primordial net-baryon stopping and of antibaryon annihilation
(subsequent to the initial antibaryon formation phase) \cite{b15,b16,b17,b18} 
the observation at top AGS energies, of $\bar{\Lambda}/\bar{p}$ significantly 
exceeding unity, remains a puzzle.
This effect might be related to the recent observation of a steep maximum of 
the $K^{+}/\pi^{+}$ ratio, occuring in the vicinity of top AGS to lowest SPS 
energy \cite{b19,b20}. Similar to the $\bar{\Lambda}/\bar{p}$ ratio, 
the $K^{+}/\pi^{+}$ ratio also refers to the anti-strange to anti-light 
quark abundance ratio, prevaling at hadron formation.

The paper presents an extension of the previously known $\bar{p}/p$
production ratio to the lower CERN SPS energies. 
New measurements of antiproton yields in centrality selected Pb+Pb collisions 
at beam energies of 158$A$ GeV ($\sqrt{s_{_{NN}}}$=17.3 GeV) and in central Pb+Pb 
collisions at 20, 30, 40 and 80 $A$ GeV ($\sqrt{s_{_{NN}}}$=6.3, 7.6, 8.7
and 12.3 GeV, respectively) were performed 
bridging the gap between the data at top AGS and SPS energies. 
Previously published proton yield measurements of NA49 at 40, 80 and 
158 $A$ GeV \cite{b21} are extended to 20 and 30 $A$ GeV beam energies. 
The $\bar{\Lambda}/\bar{p}$ ratios are then obtained using recent $\bar{\Lambda}$ 
results from NA49 \cite{b22} and NA57 \cite{b23}.

\section{Experimental method}

NA49 is a fixed target experiment at CERN with a large acceptance 
detector \cite{b24} using external SPS beams of nuclei and hadrons. The produced 
charged particles were detected in four large volume Time Projection Chambers 
(TPCs). 
Two of them (VTPCs) are located inside the magnetic field, the two others 
(MTPCs) downstream of the magnets on either side of the beam line. 
The TPCs provide precise tracking and particle identification in a wide 
range of phase space based on the measured momentum 
and the specific energy loss $dE/dx$ in the TPC gas with about 4\% resolution.
  
 Two TOF detector walls containing 891 scintillator pixels each are 
situated symmetrically behind the TPCs on both sides of the beam.  
The average overall time resolution of these detectors is $60-70$ ps.
In the present analysis, the identification of antiprotons was primarily 
accomplished by the TOF measurement, the $dE/dx$ information from the large 
TPCs being employed to reduce the background of charged pions 
and kaons in the mass spectrum. 
To demonstrate the identification capability, a sample of the particle mass 
spectra obtained from measured momenta and time-of-flight are shown in Fig.~1.

On-line event characterization and triggering is accomplished by beam 
definition detectors located in the beam line upstream of the target and 
interaction counters, and a calorimeter downstream of the target. 
The data samples were recorded with two trigger settings, providing the 
selection of central and minimum bias events.

Central Pb+Pb collisions were selected by requesting the energy deposited in 
the projectile fragmentation region to be lower than a given threshold.
This was achieved with a Zero Degree Calorimeter (ZDC) located downstream of 
the detector measuring the energy $E_{ZDC}$ of the remaining projectile 
fragments and spectator protons and neutrons.                
The upper limit on the energy in the ZDC was set to accept the 12\% 
most central events at 158$A$ GeV and 7\% at 20, 30, 40 and 80 $A$ GeV from all 
inelastic Pb+Pb collisions. 
For the top SPS energy, the 5\%
most central interactions were selected offline.

A Gas Cherenkov counter provides a minimum bias trigger for Pb+Pb collisions. 
It is placed in the gas region immediately behind the target to veto 
non-interacting projectiles. 
Triggering is accomplished by placing an upper threshold on the signal from 
this Cherenkov counter in coincidence with a valid signal from the beam 
detectors. 
Additionally, off-line cuts were made on the position of the fitted primary vertex 
along the beam direction to minimize the fraction of non-target background events.

\begin{figure}
\hspace*{0.1cm}\epsfig{file= 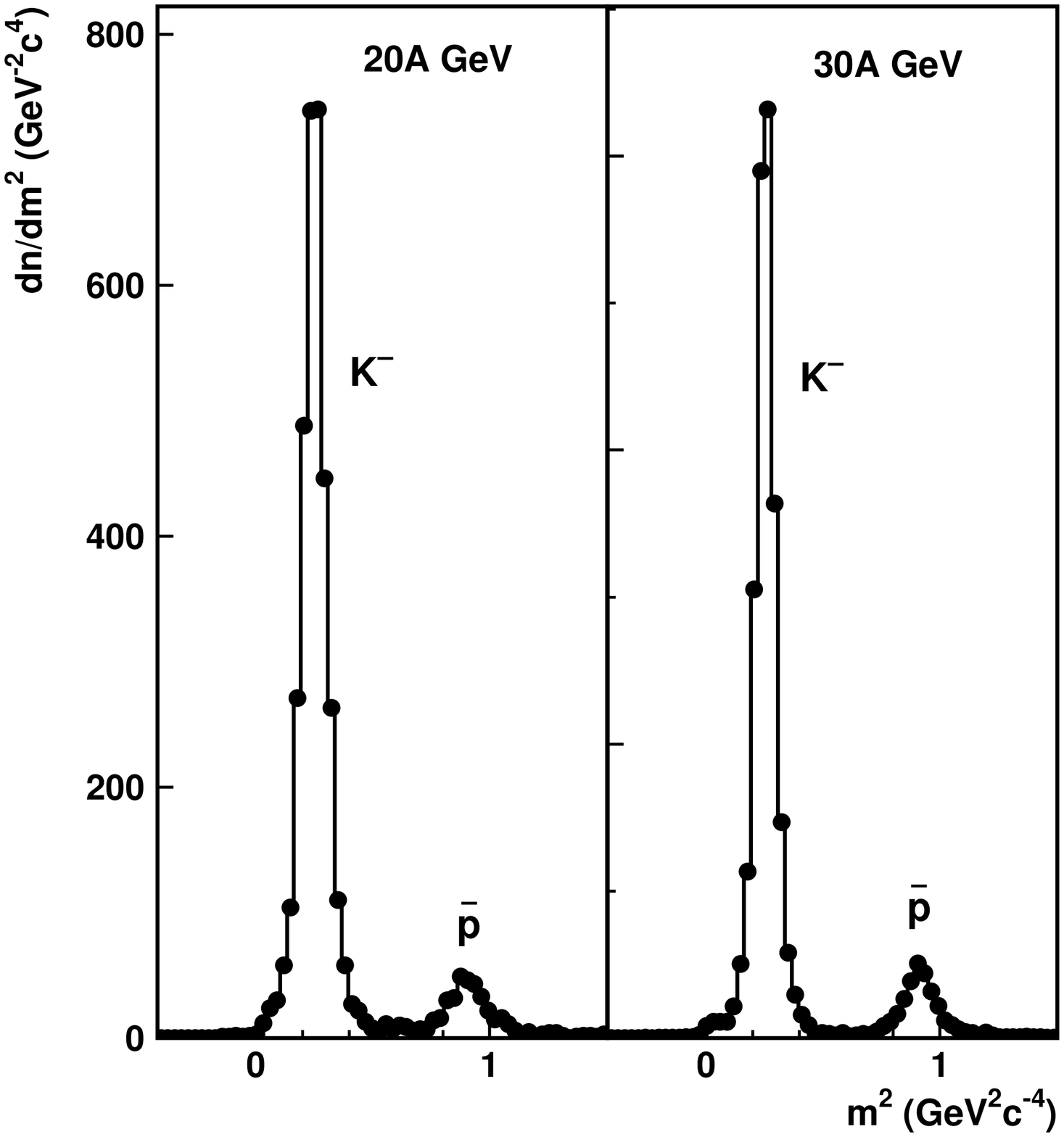 , width=7cm}
\hspace*{0.1cm}\epsfig{file= 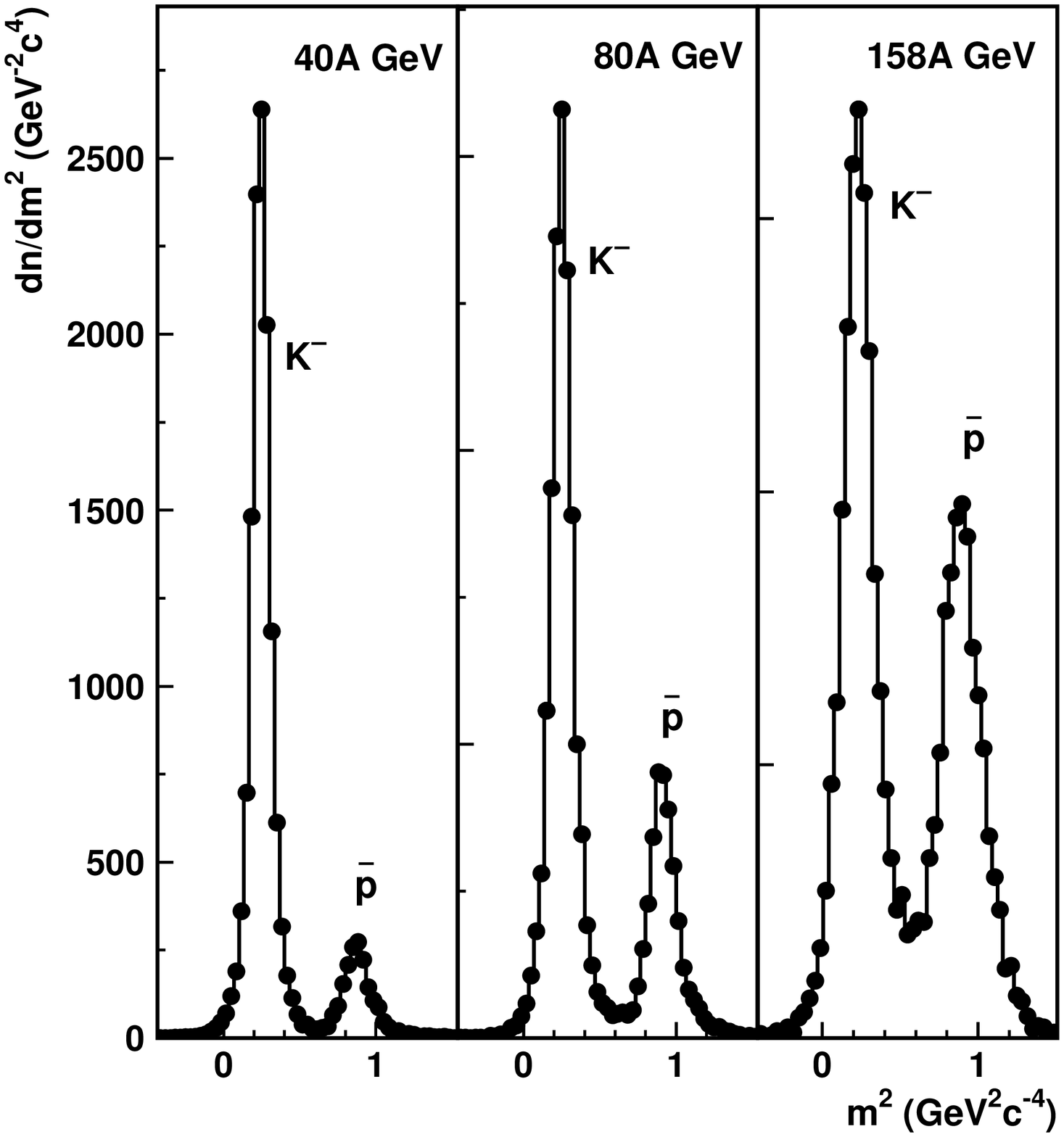 , width=7cm}
\caption{ Mass squared histograms for TOF-identified kaons and antiprotons in 
       Pb+Pb reactions at 20, 30, 40, 80 and 158 $A$ GeV beam energies.
       $dE/dx$ cuts have been applied to the data to reject most of the pions. 
       The particle momenta range from 3 to 10 GeV/$c$.  
           }
\label{fig1}
\end{figure}

\section{Data analysis and results}
 For the analysis of 158$A$ GeV Pb+Pb collisions the data samples of 450,000 
central and 400,000 minimum bias events were used. 
In order to study the centrality dependence the data were segregated into
six centrality classes, which were selected by subdividing the range of the
total energy measured in the ZDC. 
For each centrality bin the cross section is determined experimentally and 
corresponds to an interval in the inelstic Pb+Pb collision cross section. 
The resulting intervals, assuming a total inelastic cross section 
$\sigma_{inel}$ = 7.15 barns, are listed in Table I. 
The initial geometry of nucleus-nucleus collisions is usually characterized 
by the number of wounded nucleons $N_{wound}$ which essentially represent the 
number of nucleons in the geometrical overlap volume of the colliding nuclei. 
The values of $N_{wound}$ as a function of impact parameter $b$ and inelastic 
cross section range are obtained from Glauber model calculations \cite{b25}.
The resulting averages $\langle N_{wound}\rangle$ and $\langle b\rangle$
for the six inelastic cross section intervals of this analysis are shown 
in Table I. 
The value of $\sigma_{inel}$ is not known precisely.
An estimated uncertainty of 5\% leads to a systematic uncertainty in 
$\langle N_{wound}\rangle$ of 1\% ( 10\% ) for central (peripheral) collisions.
Note that bin 6 suffers from a trigger bias, i.e. the minimum bias trigger only 
accepts about 35 \% of the events in this centrality range. 
The listed values of $\langle N_{wound}\rangle$ and $\langle N_{part}\rangle$ 
are not averages for the 43 - 100\% cross section interval, but refer to those 
events accepted by the trigger. 
Since the quantity $N_{wound}$ is not a direct observable, previous publications 
of NA49 also employed the number of participants $N_{part}$, which counts all 
the nucleons from the incident nuclei involved in any hadronic reaction 
(including those hit by a scattered or produced particle).
$N_{part}$ can be estimated from the baryons and antibaryons measured by the 
detector or from the spectator energy deposited in the ZDC \cite{b26,b27,b28}. 
For comparison $\langle N_{part}\rangle$ is also listed in Table I; the values 
differ significantly from  $\langle N_{wound}\rangle$ for peripheral collisions.

\vspace{0.5cm}
\noindent{TABLE I.  
The six centrality classes used in the analysis for 158$A$ GeV Pb+Pb collisions,
listing the covered range in fraction of the total cross section $\sigma_{inel}$, 
the mean numbers of participating $\langle N_{part}\rangle$ and wounded 
$\langle N_{wound}\rangle$ nucleons, and impact parameter $\langle b\rangle$ for 
the corresponding cross sections. 
The standard deviations of the distributions of $N_{wound}$ and $b$ for the 
respective centrality bins are also quoted.}
\begin{center} 
\begin{tabular}{ c c c c c } \hline \hline 
{} &{} &{} &{} &{}\\[-1.5ex]
Centrality & Fraction~ & $\langle N_{part}\rangle$~~~ & $\langle N_{wound}\rangle$
~~ & $\langle b\rangle$  \\
 bin    & of $\sigma_{inel} (\%)$ &         &  &  (fm)   \\ \hline 
{}  &{} &{} &{} &{}\\[-1.5ex]
1   & 0-5    & 366     &   $357\pm 22$     &  $2.3\pm 0.9$ \\[1ex]
2   & 5-12   & 309     &   $288\pm 28$     &  $4.5\pm 0.8$ \\[1ex]
3   & 12-23  & 242     &   $211\pm 30$     &  $6.4\pm 0.8$ \\[1ex]
4   & 23-33  & 178     &   $146\pm 25$     &  $8.1\pm 0.8$ \\[1ex]
5   & 33-43  & 132     &   $99\pm 22$      &  $9.5\pm 0.8$ \\[1ex]
6   & 43-100 & 85      &   $42\pm 16$      &  $11.8\pm1.0$ \\[1ex] \hline \hline
\end{tabular}
\end{center}

For the reconstruction of charged tracks, a global tracking scheme was used 
which combined all track segments 
reconstructed in the TPCs that belong to the same particle.    
For each event the primary vertex was determined from the intersection 
of reconstructed tracks. Events in which no primary vertex was found 
were rejected. 
Event vertex as well as track quality cuts were applied in order to 
select the events for further analysis.
The event vertex had to lie within $\pm{1}$ cm of the target foil position. 
The global track was required to comprise track segments in the MTPC and at 
least one of the VTPCs.
The reconstructed tracks having momentum and $dE/dx$ information were 
extrapolated to the TOF detector wall and assigned a mass squared value 
$m^2$ derived from the momentum, flight path and time-of-flight measurements.
Furthermore, cuts were applied to eliminate tracks which impinge close to 
the edges of the scintillator tiles or which point to tiles hit by more than 
one track or which show a signal from 
event-correlated $\gamma$-conversion pairs in the scintillator. 
The corresponding inefficiency was determined experimentally from the TPC 
tracking data (edge and double hit) and the charge measurements in the TOF 
scintillators ($\gamma$ conversion) and amounts to 25\% 
on average, with a maximum of 30\% 
in the region of the TOF wall closest to the beam. 
The relevant corrections have been applied to the measured particle yields.

Next, antiprotons were selected and their transverse mass $m_t$ distributions 
reconstructed. The tracks were subjected to
identification cuts applied to the measured $dE/dx$ and $m^2$ values in 
order to simultaneously maximize the number of antiprotons and minimize the 
background stemming mostly from pions and kaons. 
Correction factors for the $\bar{p}$ yield due to the cuts were estimated from 
the experimental $dE/dx$ and 
$m^2$ distributions using parametrized descriptions obtained from a fit. 
The mass squared distributions were described by a sum of a Gaussian 
distribution for the $\bar{p}$ signal and a background
with two components: an exponentially falling contribution from the tails of 
$K^{-}$ and $\pi^{-}$ bands, and a flat random distribution due to  
misidentified particles. 
The energy loss $dE/dx$ is well described by a sum of Gaussian distributions. 
The fits were performed for each ($p_t,p$)-bin of width $\Delta{p_t}$=0.2 GeV/$c$
and $\Delta{p}$=1.0 GeV/$c$ in the full kinematic range of antiprotons 
accepted by the TOF detector: $0<p_t<1.7$ GeV/c and $3.0<p<10.0$ GeV/c.
The correction factors for the antiproton loss and background contamination
due to the cuts were then determined separately for each ($p_t,p$)-bin.
They were negligible in the momentum range up to $p=6.0$ GeV/$c$ and reached 
about 15\% 
at large momenta. 

Some fraction of the measured antiprotons are the daughters of weak decays 
of strange antibaryons, mainly $\bar{\Lambda}$, including those from
electromagnetic decay of $\bar{\Sigma}^0$ which are experimentally 
indistinguishable from primary $\bar{\Lambda}$.
This so called feed-down was evaluated from a GEANT-based Monte Carlo 
simulation of $\bar{\Lambda}$ and $\bar{\Sigma}$ decays in the NA49 detector, 
including detector response simulation and reconstruction of the charged 
decay products.
This procedure takes into account the probability for a secondary 
antiproton to be reconstructed as a track from the primary vertex. 
As input for the simulation, $\bar{\Lambda}$ yields and phase space 
distributions measured by the SPS experiments were employed.
The $\bar{\Lambda}$ yields in each centrality bin were 
derived using the $\bar{\Lambda}$ centrality dependence obtained in 
NA57 \cite{b23} scaled to the $\bar{\Lambda}$ yield in central Pb+Pb collisions 
measured by NA49 \cite{b22}. 
The relatively small fraction of antiprotons from $\bar{\Sigma}^-$-decays was 
calculated using the RQMD model \cite{b29} simulation.
The calculated $m_t$-spectra of feeddown antiprotons are well described 
by an exponential function.
The inverse slope parameter of these spectra gradually changes from 265 MeV 
to 215 MeV from the most central to the peripheral bins, respectively.
The overall contribution of feed-down antiprotons was found to vary from 25\% 
in peripheral to 35\% in central collisions. These values contain also the 
antiprotons from doubly strange cascades.

The correction factor for geometrical acceptance was calculated  
using the GEANT package for tracking particles and dedicated NA49 software
to simulate the detector response.
The track reconstruction efficiency in the TPCs for primary $p$ and $\bar{p}$ was 
determined by embedding simulated particle tracks into raw data events, which 
were then passed through the same reconstruction procedure as the real data. 
It was found to be nearly 100\% in the kinematical range covered by the TOF 
acceptance.

The discussed corrections have been applied to the data in each rapidity-
transverse mass ($y,m_t$)-bin. The transverse mass $m_t$ ($m_t=\sqrt{p_t^2+m^2}$) 
spectra for antiprotons were then obtained by 
integrating the data over the measured rapidity region. 
The resulting  $\bar{p}$ spectra as well as previously published proton 
spectra \cite{b21} in centrality selected 158$A$ GeV Pb+Pb collisions at 
$2.4<y<2.8$ 
($y_{cm}=2.9$) are shown in Fig.~2 along with a fit function of the form:

\begin{equation}
\frac{d^2n}{m_t dm_t dy } = C_1e^{-(m_t-m)/T} + C_2e^{-(m_t-m)/T'},
\end{equation}

A single exponential (first term in Eq.~(1)) well describes the data at 
$m_t-m>0.2$ GeV/$c^2$ and determines the inverse slope parameter of the spectra 
to be discussed further (hereafter denoted by $T$). 
The second exponential term with a slope paremeter $T'\approx{100}$ MeV accounts 
for the flattening of the spectra 
 at low $m_t$.
Its contribution to the total yield decreases from nearly 10\%
in central events to almost zero in the peripheral bins.
The double exponential parameterization of Eq.~(1) was used for extrapolation
to the unmeasured $m_t$ region when calculating the midrapidity yield $dn/dy$ 
and the mean transverse mass $\langle m_{t}\rangle -m$.
The extrapolation of the antiproton spectra ranges from 8\% 
to 15\% 
for the two most central bins and for the more peripheral bins, respectively.
Table II summarizes the results on $dn/dy$, $T$ and $\langle m_{t}\rangle -m$ 
derived from the $\bar{p}$ and $p$ transverse spectra for each centrality bin of 
158$A$ GeV Pb+Pb collisions.

\begin{figure}
\hspace*{0.1cm}\epsfig{file= 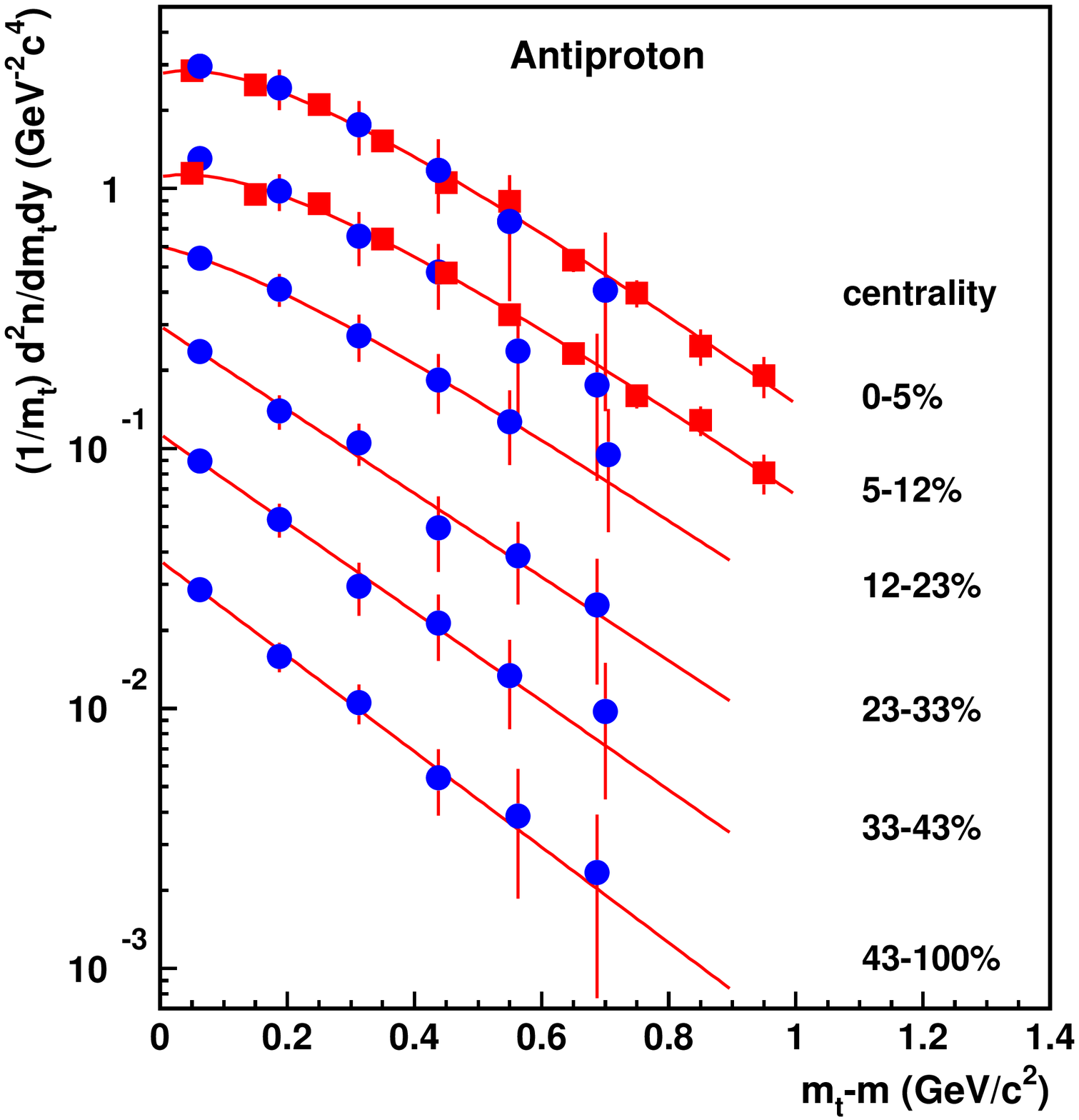 ,width=8cm}
\hspace*{0.1cm}\epsfig{file= 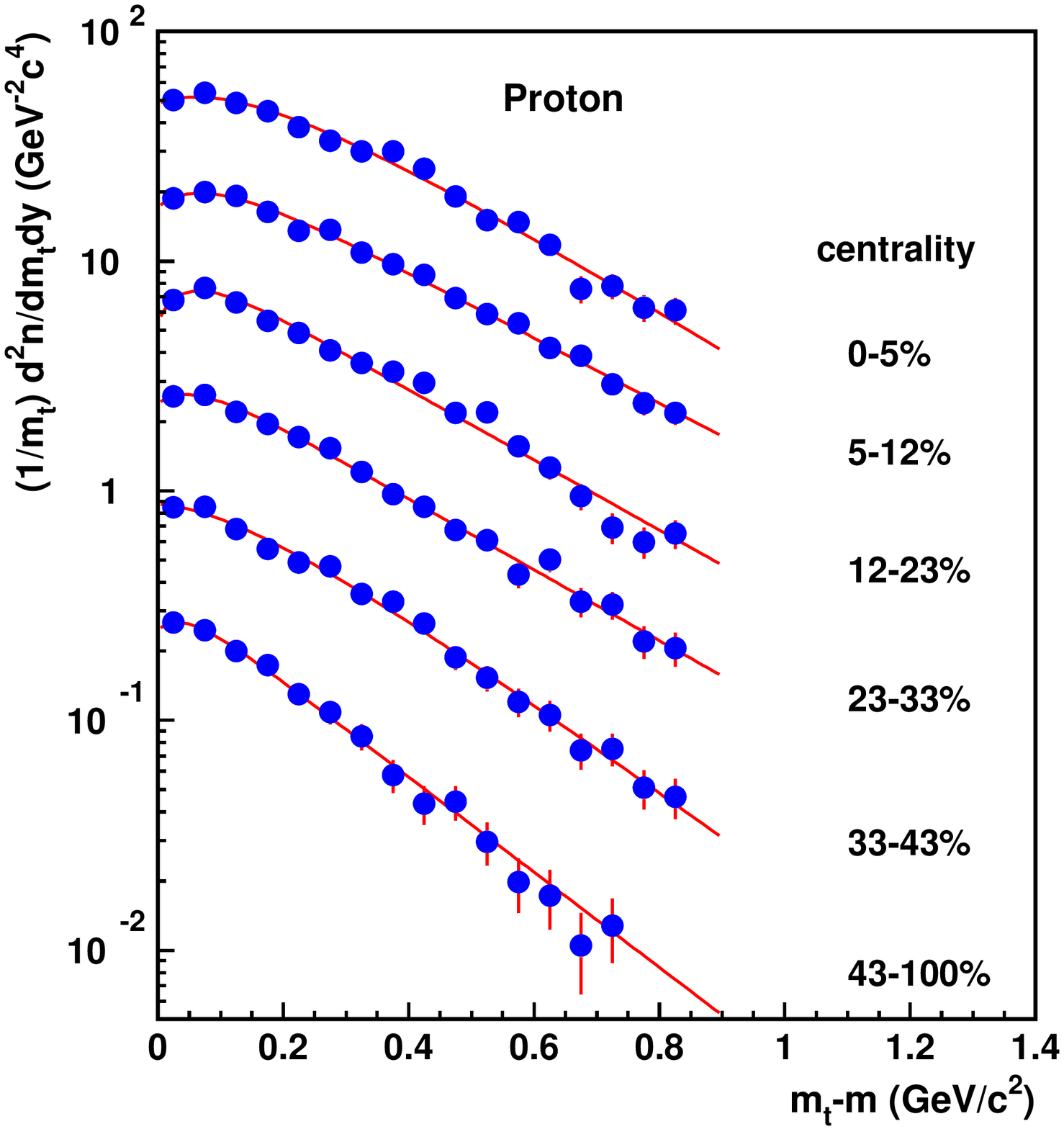 ,width=8cm}
\caption{(Color online)
Transverse mass distributions for antiprotons and protons \protect\cite{b21} 
in the rapidity interval $2.4<y<2.8$ for the six centrality classes 
in 158$A$ GeV Pb+Pb collisions. 
The circle and square points denote the data obtained with the trigger settings 
selecting minimum bias and central collisions, respectively.
The errors are statistical. 
The curves through the measured points represent the two-exponential fit of 
Eq.~(1) to the data.
For clarity, the spectra are scaled down by a factor of 2 successively from 
the uppermost data. 
           }
\label{fig2}
\end{figure}

The sources of considered systematic uncertainties include possible errors in 
the determination of the efficiency corrections, in particular those for 
selection of single track hits in the TOF pixels as well as the corrections for 
particle identification and background subtraction. This is only relevant to 
the antiprotons at higher momenta and high particle multiplicity.  

The estimate of the particle yield $dn/dy$ represents an extrapolation into
unmeasured regions under the explicit assumption of a certain shape of the $m_t$ 
distribution. However, the major systematic error arises from uncertainty
in the phase space distributions of the parent hyperons decaying into antiprotons.
The latter was derived from the spread of the experimental data on $\bar{\Lambda}$
production used for feed-down corrections including the quoted systematic errors.
The overall systematic errors of the antiproton results amount to approximately 10\%
for the particle yield and 5\% for the inverse slope parameter with a weak
dependence on centrality of the collision.

Analyses of transverse spectra at the lower SPS energies of 20, 30, 40 and 
80 $A$ GeV were performed in the same manner as for 158$A$ GeV. 
Here, the online trigger was set to the 7\% most central Pb+Pb collisions 
which corresponds to the number of wounded 
nucleons $\langle N_{wound}\rangle$=$345\pm 28$.
The data set comprises about 300,000 events for each energy.
Measurements were done near midrapidity and covered the rapidity intervals of
~$1.5<y<2.2$ for 20$A$ GeV ($y_{cm}=1.88$), ~~$1.6<y<2.3$ for 30$A$ GeV 
($y_{cm}=2.08$), ~~$1.9<y<2.3$ for 40$A$ GeV ($y_{cm}=2.22$) and ~~$2.2<y<2.6$ 
for 80$A$ GeV ($y_{cm}=2.57$).

\vspace{0.3cm}
\noindent{ TABLE II.
 Particle yield $dn/dy$, inverse slope $T$ and
mean transverse mass $\langle m_{t}\rangle -m$ for antiprotons and 
protons ($2.4<y<2.8$) at various centralities in 158$A$ GeV Pb+Pb collisions. 
The errors are statistical. Proton results are from \protect\cite{b21}.}
\begin{center}
\begin{tabular}{c c c c c}\hline\hline
{} &{} &{} &{} &{}\\[-1.5ex]
~~~~~~~~ & Centrality  &  $dn/dy$         &  $T$    & $\langle m_{t}\rangle -m$ \\
           & (\% of $\sigma_{inel}$) &          & (MeV)   &  (MeV/$c^2$)     \\ \hline
{} &{} &{} &{} &{}\\[-1.5ex]
$\bar{p}$ & ~~0-5 & ~$1.66\pm 0.17$  &  ~~~$291\pm 15$  & ~~~$384\pm 19$ \\[1ex]
        & ~~5-12   & ~$1.27\pm 0.11$ &  ~~~$299\pm 15$  & ~~~$393\pm 16$ \\[1ex]
        & ~~12-23  & ~$1.05\pm 0.08$ &  ~~~$274\pm 22$  & ~~~$370\pm 45$ \\[1ex]
        & ~~23-33  & ~$0.76\pm 0.06$ &  ~~~$269\pm 29$  & ~~~$320\pm 35$ \\[1ex]
        & ~~33-43  & ~$0.55\pm 0.05$ &  ~~~$255\pm 28$  & ~~~$309\pm 32$ \\[1ex]
        & ~~43-100 & ~$0.33\pm 0.04$ &  ~~~$236\pm 26$  & ~~~$284\pm 28$ \\[2ex]
{} &{} &{} &{} &{}\\[-1.5ex] 
$p$      & ~~0-5    & ~$29.6\pm 0.9$ &  ~~~$308\pm 9$   & ~~~$413\pm 13$ \\[1ex]
        & ~~5-12   & ~$22.2\pm 0.6$  &  ~~~$308\pm 9$   & ~~~$415\pm 14$ \\[1ex]
        & ~~12-23  & ~$14.5\pm 0.4$  &  ~~~$276\pm 9$   & ~~~$362\pm 12$ \\[1ex]
        & ~~23-33  & ~$9.8\pm 0.3$   &  ~~~$273\pm 10$  & ~~~$355\pm 12$ \\[1ex]
        & ~~33-43  & ~$5.7\pm 0.2$   &  ~~~$245\pm 10$  & ~~~$315\pm 13$ \\[1ex]
        & ~~43-100 & ~$2.9\pm 0.1$   &  ~~~$216\pm 10$  & ~~~$259\pm 12$ \\[1ex] \hline\hline
\end{tabular}
\end{center}

\begin{figure}
\hspace*{0.1cm}\epsfig{file= 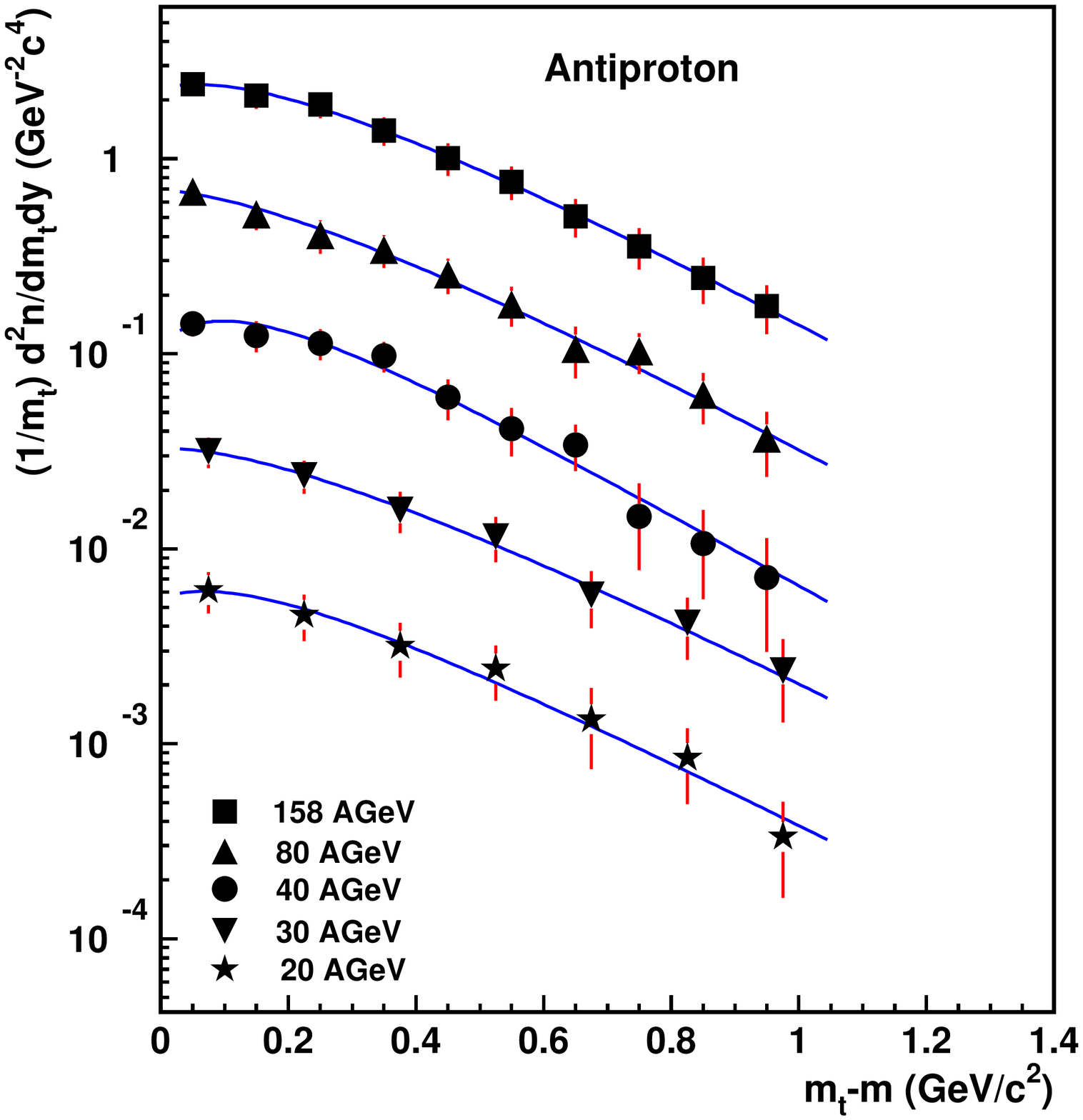 ,width=8cm}
\hspace*{0.1cm}\epsfig{file= 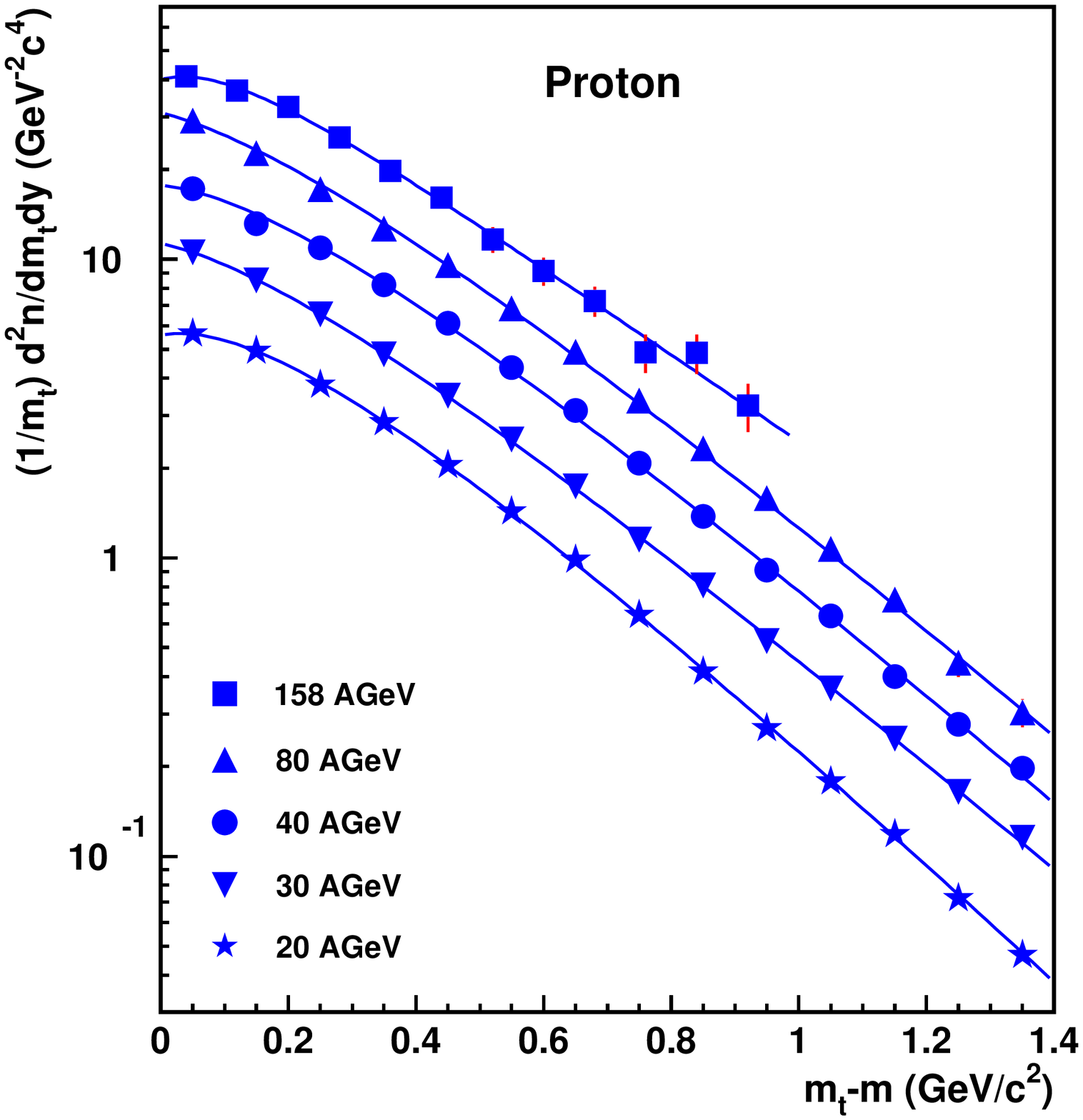 ,width=8cm}
\vspace{0.3cm}
\caption{(Color online)
Transverse mass distributions for antiprotons and protons at midrapidity in
central Pb+Pb collisions at 20, 30, 40, 80 and 158 $A$ GeV. 
The errors are statistical.
The solid lines illustrate the two-exponential fit of Eq.~(1) to the data.
For clarity, the spectra are scaled down by a factor of 2 successively from 
the uppermost data. 
           }
\label{fig3}
\end{figure}

The geometrical acceptance at lower energies allows to extend the transverse 
mass distributions for protons up to $m_t-m=1.5$ GeV/$c^2$ ($p_t=2.25$ GeV/$c$). 
Thus, the extrapolation into the unmeasured $m_t$ region is negligibly small.  
The background contamination due to misidentification of antiprotons at 
lower energies is less than that at 158$A$ GeV.

Corrections due to the identification cuts applied to $dE/dx$ and $m^2$ for
selection of the $\bar{p}$ samples do not exceed a few percent at the largest 
values of $p$ and $p_t$. For the feed-down correction, 
the NA49 data \cite{b22,b30} on $\bar{\Lambda}$ and $\Lambda$  production 
obtained for the same data samples were used.
The $\bar{p}$ feed-down contribution at lower energies became larger as
compared to 158$A$ GeV and gradually increases with decreasing beam energy 
reaching nearly 50\% at 20 and 30 $A$ GeV.
This increases the overall systematic errors towards lower energies
approaching 15\% for the antiproton yield at 20$A$ GeV.
The feed-down correction for $p$ at low energies amounts to approximately
15\% which is rather close to that for 158$A$ GeV.

Fig.~3 depicts the $\bar{p}$ spectra for central Pb+Pb collisions at all five
energies together with fits by the double exponential function Eq.~(1).
A deviation from a single exponential shape is clearly seen. 
The slope parameter $T'\approx{100}$ MeV in the second term of Eq.(1) is almost 
the same for all five energies. The contibution of this term to the total yield 
is estimated to be about 10\%.
The numerical values for $dn/dy$, $T$ and $\langle m_{t}\rangle -m$ are listed
in Table III.

\vspace{0.3cm}
\noindent{ TABLE III. Particle yield $dn/dy$, inverse slope $T$ and
mean transverse mass $\langle m_{t}\rangle -m$ for antiprotons and 
protons in central Pb+Pb collisions at 20, 30, 40, 80 and 158 $A$ GeV
beam energies. 
The errors are statistical.
                         }
\begin{center}
\begin{tabular}{c c c c c}\hline\hline
{} &{} &{} &{} &{}\\[-1.5ex]
~~~~~~~~   &  $E_{beam}$ &   $dn/dy$  &  $ T $   &  $\langle m_{t}\rangle-m$  \\
           &  ($A$GeV)  &            &  (MeV)   &     (MeV/$c^2$) \\ \hline
{} &{} &{} &{} &{}\\[-1.5ex]
 $\bar{p}$ & ~~~~158  & ~~~$1.66\pm 0.17$  &  ~~~~$291\pm 15$  & ~~~~$384\pm 19$ \\[1ex]
           & ~~~~80   & ~~~$0.87\pm 0.07$  &  ~~~~$283\pm 30$  & ~~~~$385\pm 41$ \\[1ex]
           & ~~~~40   & ~~~$0.32\pm 0.03$  &  ~~~~$246\pm 35$  & ~~~~$355\pm 51$ \\[1ex]
           & ~~~~30   & ~~~$0.16\pm 0.02$  &  ~~~~$290\pm 45$  & ~~~~$395\pm 60$ \\[1ex]
           & ~~~~20   & ~~~$0.06\pm 0.01$  &  ~~~~$279\pm 64$  & ~~~~$394\pm 60$ \\[2ex]
{} &{} &{} &{} &{}\\[-1.5ex]
 $p$       & ~~~~158  & ~~~$29.6\pm 0.9$   & ~~~~$308\pm  9$   & ~~~~$413\pm 13$ \\[1ex]
           & ~~~~80   & ~~~$30.1\pm 1.0$   & ~~~~$260\pm 11$   & ~~~~$364\pm 16$ \\[1ex]
           & ~~~~40   & ~~~$41.3\pm 1.1$   & ~~~~$257\pm 11$   & ~~~~$367\pm 16$ \\[1ex]
           & ~~~~30   & ~~~$42.1\pm 2.0$   & ~~~~$265\pm 10$   & ~~~~$362\pm 14$ \\[1ex]
           & ~~~~20   & ~~~$46.1\pm 2.1$   & ~~~~$249\pm  9$   & ~~~~$352\pm 13$ \\[1ex]\hline\hline
\end{tabular}
\end{center}
\vspace{0.3cm}

\section{Discussion}

The midrapidity $m_t$-spectra for 158$A$ GeV Pb+Pb collisions become progressively 
flatter from peripheral to central collisions (Fig.~2 and Table II). Both the mean 
transverse mass $\langle m_{t}\rangle-m$ and the inverse slope parameter $T$ 
increase towards central collisions, although this trend is slightly different 
for $\bar{p}$ and $p$, namely the proton inverse slope increases somewhat faster, 
as clearly seen in Fig.~4. 

\begin{figure}
\hspace*{-0.3cm}\epsfig{file= 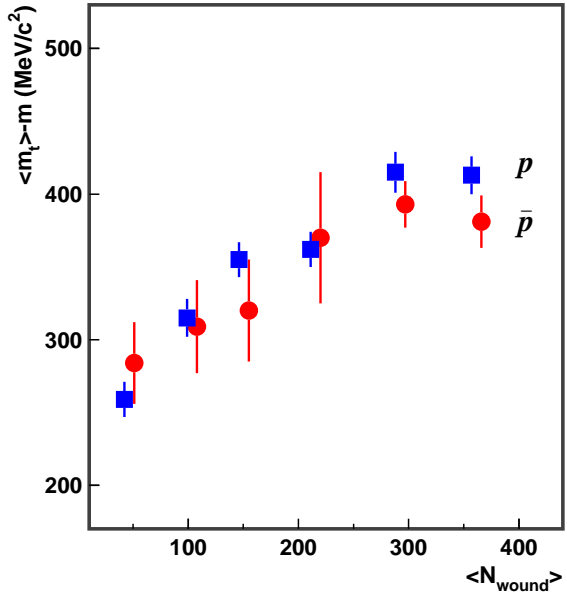 ,width=9cm}
\caption{(Color online)
 The mean transverse mass $\langle m_{t}\rangle-m$ for $\bar{p}$ and $p$ as a function of 
 the number of wounded nucleons $N_{wound}$ at midrapidity in 158$A$ GeV Pb+Pb 
 collisions.
The errors are statistical.  
           }         
\label{fig4}
\end{figure}

A deviation from a single exponential at low $m_t$ values, the so called 
"shoulder-arm shape", is most pronounced in central Pb+Pb collisions (Fig.~2) 
and observed at all five energies (Fig.~3).
These features are generally explained by a strong collective expansion 
(radial flow) in hydrodynamical models \cite{b31,b32,b33}.
A blast wave parameterization based on these models provides in fact as good
a description of the $m_t$-spectra \cite{b34} as the double exponential
function. Fitted parameters yield an average radial flow velocity of about
0.5 c and a temperature of 120-140 MeV in the expanding fireball at its
disintegration stage.

Interestingly, the $\bar{p}$ and $p$ transverse mass spectra reveal within 
errors quite similar shapes in spite of the fact that at these energies a 
significant fraction of protons carry baryon number from the incoming nuclei, 
while antiprotons are predominantly pair produced. This observation may indicate 
similar expansion dynamics for both particles comprised of thermal 
motion and collective radial flow that determine the transverse 
spectra of particles at the late stage of the expansion. 
As discussed in the introduction, substantial annihilation of primordially 
produced $\bar{p}$ might occur in the baryon rich fireball. This effect is
expected to lead to a decrease of the $\bar{p}$/$p$ ratio at $m_t-m$ below 
about 0.4 GeV/$c^2$ \cite{b15,b17}. In particular, at 158$A$ GeV the close
similarity of the shapes of the measured $\bar{p}$ and $p$ spectra do not support 
such a prediction. At lower energies a decrease of the order of 30\%
cannot be ruled out due to the larger uncertainties of the measurements. 

In contrast to the strong centrality dependence, the shape of the measured 
transverse mass spectra does not change noticeably with beam energy 
at SPS energies. This is illustrated in Fig.~5 where the $\langle m_{t}\rangle-m$ 
values for  $p$ and $\bar{p}$ in central Pb+Pb collisions at the five SPS energies 
are plotted versus the nucleon-nucleon center-of-mass energy. Also shown 
are measurements at lower and higher energies from the AGS \cite{b35,b36,b37} 
and RHIC \cite{b38}, respectively. 
It is seen that the mean transverse mass for both $\bar{p}$ and $p$ steeply rises 
with energy at the AGS, remains approximately constant at the SPS and resumes 
a slow rise towards RHIC energies. 
A similar non trivial energy dependence was also observed by NA49 for kaon 
and pion production \cite{b30} which was considered a possible manifestation of 
the coexistence of partonic and hadronic phases or more generally a very soft 
equation of state in the SPS energy regime \cite{b39,b40,b41}. The behavior of 
the $m_t$ spectra may thus be taken to support the indication for an onset of 
deconfinement in the early stage of the collisions at low SPS energies as 
obtained from the energy dependence of pion and kaon yields \cite{b19}.

\begin{figure}

\hspace*{0.0cm}\epsfig{file= 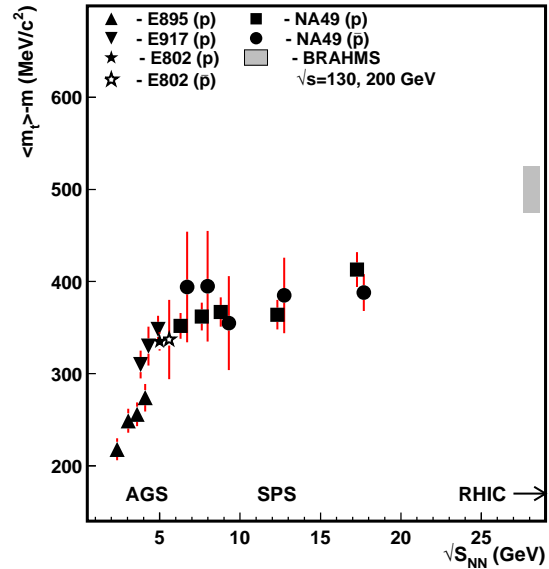 ,width=8cm}
\vspace{-0.5cm}
\caption{(Color online)
The mean transverse mass $\langle m_{t}\rangle-m$ for $\bar{p}$ and $p$ at 
midrapidity as a function of the center-of-mass energy per nucleon pair 
$\sqrt{s_{_{NN}}}$ in central Pb+Pb collisions at SPS energies (NA49) together 
with data for central Au+Au collisions from the AGS 
\protect\cite{b35,b36,b37} and RHIC \protect\cite{b38}.  
The errors are statistical.
           }         
\label{fig5}
\end{figure}

The measured particle yields $dn/dy$ differ significantly between $p$ and $\bar{p}$. 
Both increase with increasing centrality, but the proton yield rises about 
two times faster. 
The yields of $p$ and $\bar{p}$ per wounded nucleon 
$(dn/dy)/\langle N_{wound}\rangle$ as a function of $N_{wound}$ for the six centrality 
samples are depicted in Fig.~6. 
An increase of this ratio with centrality for protons is seen which can be understood as 
a consequence of the 
increased baryon stopping in central Pb+Pb collisions \cite{b26,b27,b28}. 
On the other hand, the $\bar{p}$ yield per wounded nucleon shows a decrease with 
increasing centrality. 
This could be a sign of increasing antiproton absorption in central collisions.
This effect was studied with the microscopic models UrQMD [17] and HSD [5] which 
comprise string excitation, hadronisation and rescattering (transport) stages. 
The UrQMD results [17] suggest a concurrence of the $\bar{p}$ enhancement in Pb+Pb 
collisions with respect to p+p reactions and an increasing antibaryon absorption 
with centrality due to the growth of the net baryon density at midrapidity.
However, the observation of nearly identical spectra for $p$ and $\bar{p}$ does not 
support the UrQMD model calculation.
The HSD model explains the weak centrality dependence of the $\bar{p}$
yield per wounded nucleon by the contribution of multi-meson reactions
in the transport phase [5]. These regenerate $\bar{p}$ and thus counteract
the loss by absorption.

\begin{figure}
\hspace*{0.1cm}\epsfig{file= 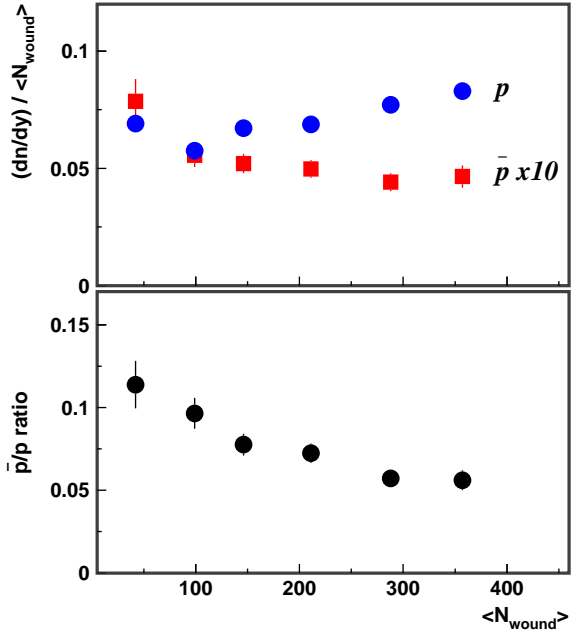 ,width=8.5cm}
\caption{(Color online)
The yield $dn/dy$ of $\bar{p}$ and $p$ per number of wounded nucleons $N_{wound}$
and the $\bar{p}/p$ ratio plotted as a function of $N_{wound}$ at midrapidity 
in 158$A$ GeV Pb+Pb collisions.
The errors are statistical.  
             }         
\label{fig6}
\end{figure}

The different dependence of $\bar{p}$ and $p$ yields on $N_{wound}$ 
results in significant centrality dependence of the $\bar{p}/p$ ratio as 
shown in the bottom panel of Fig.~6.
This ratio steadily decreases with centrality by a factor of about two within 
the considered centrality range.
A similar result was obtained in Au+Au collisions at 11.7$A$ GeV \cite{b37}.   
The data for smaller collision systems indicate an even larger pbar/p ratio 
than observed in peripheral Pb+Pb reactions reaching values of nearly 0.3 for 
p+p \cite{b42} and p+Be, p+S and p+Pb \cite{b43} interactions.
In contrast the $\bar{p}/p$ ratio measured at RHIC shows very weak variation
with centrality at the collision energy $\sqrt{s_{_{NN}}}$=130 GeV \cite{b44,b45}
and almost no dependence on centrality at 
$\sqrt{s_{_{NN}}}$=200 GeV \cite{b46,b47}.
Thus, the net baryon density, which is significant at lower energies, 
strongly affects the midrapidity $p$ and $\bar{p}$ abundances through baryon 
stopping and baryon-antibaryon annihilation in central collisions at the AGS 
and SPS, but not at RHIC energies. 

\begin{figure}
\hspace*{0.1cm}\epsfig{file= 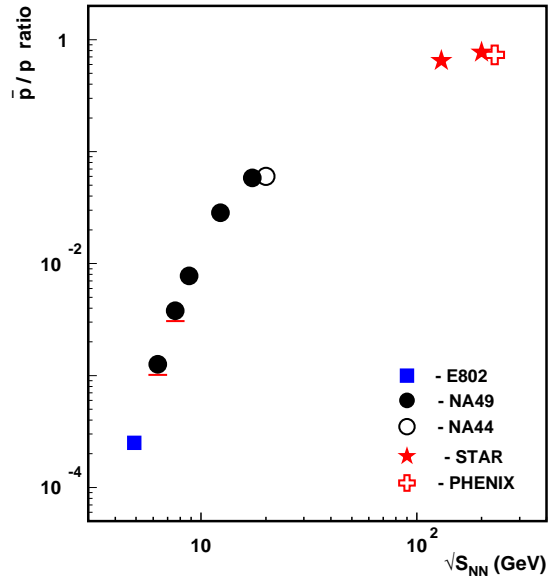 ,width=8cm}
\caption{(Color online)
The $\bar{p}/p$ ratio at midrapidity as a function of the center-of-mass 
energy per nucleon pair $\sqrt{s_{_{NN}}}$ in central Pb+Pb collisions at SPS 
energies (NA49) together with the data for lower energies at the AGS 
\protect\cite{b37} and higher energies at the RHIC 
\protect\cite{b44,b45,b47}, respectively. 
The data \protect\cite{b48} at top SPS energy are also shown.
           } 
\label{fig7}
\end{figure}

Fig.~7 displays the midrapidity $\bar{p}/p$ ratio for central collisions 
as a function of energy at AGS, SPS and RHIC.
The numerical values are listed in Table IV. The $\bar{p}/p$ ratio rises 
steeply within the SPS energy range by nearly two orders of magnitude. 
The figure illustrates how the collisions evolve from producing a net 
baryon-rich system at the AGS through the SPS energy range to an almost 
net baryon-free midrapidity region at the RHIC. 

Measurements of $\bar{\Lambda}$ production in central Pb+Pb collisions 
\cite{b22,b30} make it possible to analyse strange and nonstrange 
antibaryon production for all five beam energies. Of particular interest 
is the $\bar{\Lambda}/\bar{p}$ ratio which was briefly discussed in the 
introduction.
Note, that the $\bar{\Lambda}$ yields used for calculations contain 
the contribution from electromagnetic decays of $\bar{\Sigma}^0$ hyperons, 
which are experimentally indistinguishable from those created in primary 
interactions.

The measured values of the $\bar{\Lambda}/\bar{p}$ ratio for central Pb+Pb 
collisions at 20, 30, 40, 80 and 158 $A$ GeV are listed in Table IV and 
plotted in Fig.~8 together with those from AGS and RHIC. 
The AGS experiments reported a $\bar{\Lambda}/\bar{p}$ ratio of about 3-3.5 
for central Au+Au \cite{b7,b8} and Si+Au \cite{b9} collisions at beam momenta 
of 11.7$A$ and 14.6$A$ GeV/$c$, respectively.
As illustrated in Fig.~8 the measurements at the SPS indicate a gradual 
increase of the $\bar{\Lambda}$/$\bar{p}$ ratio from 158$A$ GeV to 30 and 
20 $A$ GeV, and tend to corroborate the large values for this ratio 
found at AGS energies. 
At 158$A$ GeV the published prediction \cite{b18} for the midrapidity
$\bar{\Lambda}$/$\bar{p}$ ratio from the UrQMD model, which takes into account 
antibaryon absorption, agrees well with the measured value.
Predictions for the full energy range are not yet available in the 
literature.
Since both $\bar{\Lambda}$ and $\bar{p}$ are newly produced baryons having no 
valence quarks in common with the projectile nucleons we may compare
the midrapidity ratio with the full phase space multiplicity ratio 
predicted by the statistical hadron gas model \cite{b49} which uses a 
smooth parameterisation of the energy dependence of the baryochemical 
potential. As demonstrated by the curve in Fig.~8 the hadron gas model 
underpredicts the ratio but shows a rise towards lower energies similar 
to the measurements. Similar predictions were obtained within 
non-equilibrium versions of the hadron gas model \cite{b50,b51}.

\begin{figure}
\hspace*{-0.3cm}\epsfig{file= 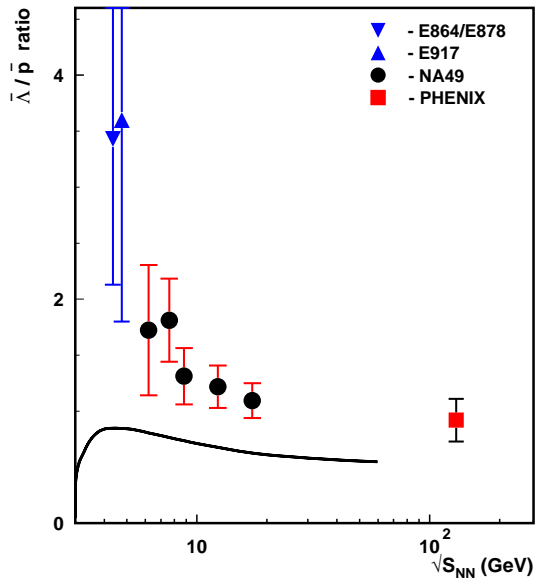 ,width=8cm}
\caption{(Color online)
The $\bar{\Lambda}/\bar{p}$ ratio at midrapidity as a function 
of the center-of-mass energy per nucleon pair $\sqrt{s_{_{NN}}}$ 
in central Pb+Pb collisions at SPS energies (NA49) together with 
the data from AGS \protect\cite{b7,b8} and RHIC \protect\cite{b11}.  
The total error bars are drawn. 
The curve shows the prediction of the statistical hadron gas model 
\protect\cite{b49}.
           } 
\label{fig8}
\end{figure}

\begin{figure}
\hspace*{-0.6cm}\epsfig{file= 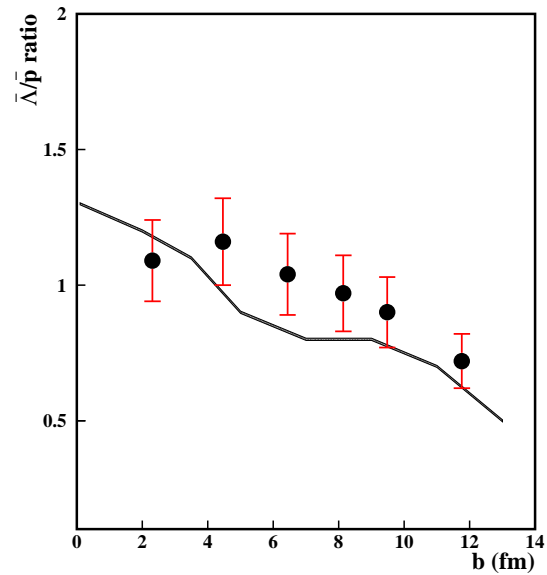 ,width=8cm}
\caption{(Color online)
Impact parameter dependence of the $\bar{\Lambda}/\bar{p}$ ratio at
midrapidity
in central Pb+Pb collisions at 158$A$ GeV. The errors are statistical.
The curved line presents the result of hadronic cascade (UrQMD)
calculation \protect\cite{b18}.
           }
\label{fig9}
\end{figure}

The increase of the $\bar{\Lambda}/\bar{p}$ ratio towards top AGS energy
may also find an explanation in a quark coalescence model scenario. Note
first of all that we deal here with an antihyperon to antiproton maximum
because this ratio has to fall down again at yet lower energies (where no
data exist due to insufficient statistics) owing to the higher
$\bar{\Lambda}$ production threshold.
Such an antihyperon maximum is reminiscent of the maximum in the same
energy range that was reported recently for the $K^{+}/\pi^{+}$ production
ratio \cite{b19,b20}.
The steep maximum in the relative strangeness production was predicted as a
signal of the onset of deconfinement \cite{b52}.
In fact if hadronization occurs by quark coalescence at the QGP phase
boundary the $\bar{\Lambda}/\bar{p}$ ratio essentially reflects the ratio of
$\bar{s}$ to $\bar{u}$ quark densities, while the $K^{+}/\pi^{+}$ ratio 
follows from the $\bar{s}$ to $\bar{d}$ ratio. 
The $\bar{d}$ quark density is expected to be proportional, in turn, to the 
$\bar{u}$ quark density.
These considerations thus provide a possible explanation of the similar rise 
of the $\bar{\Lambda}/\bar{p}$ and $K^{+}/\pi^{+}$ ratios with decreasing
collision energy.

\vspace{0.3cm}
\noindent{Table IV. The $\bar{p}/p$ and $\bar{\Lambda}/\bar{p}$ ratios at
midrapidity in central Pb+Pb collisions at SPS energies. The errors are 
statistical.
Preliminary results for the $\bar{\Lambda}$ yields at 20 and 30 $A$ GeV
\cite{b30} are used.
\vspace{-0.2cm}
\begin{center}
\begin{tabular}{c c c}\hline\hline
{} &{} &{} \\[-1.5ex]
~~~~~~~$E_{beam}$($A$GeV)    & $\bar{p}/p$ ratio~~~~~    &
$\bar{\Lambda}/\bar{p}$ ratio~~~
~~~~  \\
                     &                    &             \\ \hline
{} &{} &{} \\[-1.5ex]
  158     & $0.058\pm  0.005$~~~~~~   & $1.09\pm 0.15$~~~~~~   \\[1ex]
   80     & $0.028 \pm 0.003$~~~~~~   & $1.22\pm 0.14$~~~~~~   \\[1ex]
   40     & $0.0078\pm 0.0010$~~~~~   & $1.31\pm 0.19$~~~~~~   \\[1ex]
   30     & $0.0038\pm 0.0008$~~~~~   & $1.81\pm 0.37$~~~~~~   \\[1ex]
   20     & $0.0013\pm 0.0002$~~~~~   & $1.72\pm 0.58$~~~~~~   \\[2ex]
\hline\hline
\end{tabular}
\end{center}

For 158$A$ GeV Pb+Pb collisions the data allow a study of the centrality 
dependence of the midrapidity $\bar{\Lambda}$/$\bar{p}$ ratio which has been
suggested to be particulary sensitive to the interplay between production 
and subsequent absorption.
As data on the centrality dependence of $\bar{\Lambda}$ production are not yet 
available from NA49, numerical values for the ratio were derived from the yield 
of $\bar{\Lambda}$ in the most central bin 1 measured by NA49 \cite{b22} and the 
calculated yields for each centrality bin 2-6 using the NA57 result \cite{b23} 
that the $\bar{\Lambda}$ yield within the errors is proportional to the number 
of wounded nucleons $N_{wound}$.  
As shown in Fig.~9 the $\bar{\Lambda}$/$\bar{p}$ ratio increases from peripheral 
to central collisions by about a factor 2. 
A much stronger increase with centrality has been observed at the AGS \cite{b7,b8}. 
Interestingly, the results at 158$A$ GeV are in agreement with a UrQMD model 
calculation \cite{b18} which incorporates antibaryon absorption.


\section{Summary}

The midrapidity transverse mass distributions for antiprotons 
were measured in Pb+Pb collisions at 20, 30, 40, 80 and 158 $A$ GeV 
filling the gap in the data between AGS and top SPS energies. They are 
compared to the relevant data for proton and antilambda production.

The shapes of $m_t$-spectra show a strong dependence on collision centrality. 
In central collisions and at small $m_t$ the spectra exhibit a pronounced 
deviation from the exponential Boltzmann shape. 
The data reveal the shoulder-arm structure characteristic of prominent 
radial collective expansion flow.
The observation of nearly identical spectral shape for antiprotons and protons 
indicates a similar expansion history for both particle species. 

No visible change in the mean transverse mass for $\bar{p}$ and $p$ was found
in central Pb+Pb collisions within the measured SPS energy range.
This is similar to what has been observed in kaon and pion production and could 
be attributed to the possible formation of a deconfined phase in the early stage 
of the collision.

The yield of $p$ normalized to the number of wounded nucleons increases with 
centrality while this quantity for $\bar{p}$ exhibits a decrease. The former effect 
can be understood as the result of an increase of baryon stopping with collision 
centrality which leads to contraction of the proton rapidity distribution around 
midrapidity. 
The decrease of the yield of $\bar{p}$ per wounded nucleon may point to some 
contribution from annihilation.

The $\bar{p}/p$ ratio is found to increase by almost two orders of magnitude in central 
Pb+Pb collisions at 158$A$ GeV as compared to 20$A$ GeV beam energy, reflecting the 
rapid decrease of the net baryon density at midrapidity with collision energy.

In central Pb+Pb collisions the $\bar{\Lambda}/\bar{p}$ ratio shows a steady increase 
with decreasing beam energy approaching a value of almost 2 at the energies of 20 and 
30 $A$ GeV. 
This confirms earlier evidence for this ratio to significantly exceed unity at AGS energies.
There may be an analogy to the $K^{+}/\pi^{+}$ maximum also observed in the domain 
of top AGS to lowest SPS energies. 

No fully satisfactory theoretical description of antibaryon production is
available in the literature. The new experimental results on $\bar{p}$ 
production should help to elucidate the possible role of annihilation processes 
in antihyperon production in collisions of heavy nuclei.

\section{Acknowledgments}
This work was supported by the US Department of Energy Grant DE-FG03-97ER41020/A000, 
the Bundesministerium fur Bildung und Forschung, Germany, the Virtual Institute VI-146 
of Helmholtz Gemeinschaft, Germany, the Polish State Committee for Scientific Research 
(1 P03B 097 29, 1 PO3B 121 29, 
2 P03B 04123), the Hungarian Scientific Research Foundation (T032648, T032293, T043514), 
the Hungarian National Science Foundation, OTKA, (F034707), the Polish-German Foundation, 
the Korea Research Foundation Grant (KRF-2003-070-C00015) and the Bulgarian National Science 
Fund (Ph-09/05).

\end{document}